\documentclass{appolb}
\usepackage{epsfig}

\begin{document}
\title{Overdamped Deterministic Ratchets Driven By Multifrequency Forces
\thanks{This work is dedicated to Professor Peter Talkner on the occasion of his 60th birthday.}
}
\author{David Cubero, Jes\'us Casado--Pascual, Azucena Alvarez,  Manuel Morillo
\address{Universidad de Sevilla. Facultad de F{\'{\i}}sica. Apdo. Correos
  1065.\\ Sevilla 41080. Spain}
\and
Peter H{\"a}nggi
\address{Institut f\"ur Physik,
Universit\"at Augsburg, Universit\"atsstra\ss e 1, D-86135 Augsburg,
Germany }
}
\maketitle
\begin{abstract}
We investigate a dissipative, deterministic ratchet model in the
overdamped regime driven by a {\it rectangular force}. Extensive
numerical calculations are presented in a diagram depicting the
drift velocity as a function of a wide range of the driving
parameter values. We also present some theoretical considerations
which explain some features of the mentioned diagram. In particular,
we proof the existence of regions in the driving parameter space
with bounded particle motion possessing zero current. Moreover, we
present an explicit analytical expression for the drift velocity in
the adiabatic limit.
\end{abstract}
\PACS{05.60-k, 05.45.Pq, 05.45.Ac, 05.45.Xt}

\section{Introduction and model set-up}
Directed current in ratchet systems have received much attention
over recent years
\cite{hanbar96,astumian2002,linke02,hanggi05,reimann02}. One of the
reasons to study these type of systems is motivated by the attempt
to understand the physical mechanism of motion of molecular motors
in biological systems \cite{astumian2002,prostRMP} and to
investigate its role in the design of new material properties
\cite{linke02,hanggi05}. The dynamics of a particle moving in a
periodic potential under the action of an applied time periodic term
is rather complex and rich, displaying typically even chaotic
behavior \cite{junkis96,mat00,sengupta04,ferlar05}. A feature of
particular interest is the emergence of directed currents in the
system response to an external time-periodic driving with zero
time-averaged value. Though most of the works about ratchet systems
consider the presence of noise, this phenomenon may also arise in
deterministic systems, both in the overdamped
\cite{barhan94,sarlar99,popari00} and underdamped
\cite{junkis96,mat00} regimes. Moreover, the phenomenon of
anticipated synchronization occurring in inertial deterministic
ratchets have been studied recently \cite{talkneritem}.\\

In particular, in the overdamped regime, the one-dimensional
particle dynamics $x(t)$ usually considered is governed by a first
order differential equation of the type
\begin{equation}
\dot{x}(t)=-U^{\prime}[x(t)]+F(t),
\label{eq:eom}
\end{equation}
where the dot and the prime denote time and spatial derivatives, respectively,
$U(x)$ is a periodic potential with spatial period $\lambda$ [i.e., $U(x+\lambda)=U(x)$],
and $F(t)$ is a time-periodic driving force with period $T$ [i.e., $F(t+T)=F(t)$].
In this type of systems, the current is defined as the average velocity
\begin{equation}
v=\lim_{t \to \infty} \frac {x(t)-x(0)}t.
\label{eq:veldef}
\end{equation}
As shown in Ref.~\cite{barhan94} co-existing attractors can exist
for large driving strengths, which however are not current-carrying.
A finite current $v$, possessing an unbounded  $x(t)$-trajectory, is
consequently independent of the initial condition $x(0)$. It can
also be shown that a directed current ($v\ne 0$) is only possible if
at least one of the following symmetries is broken
\cite{astumian2002,hanggi05,reimann02,adjmuk94}:
\begin{equation}
\exists x_0\in\Re \quad \mbox{such that}\quad U(x_0-x)=U(x_0+x) \quad \forall x\in\Re,
\label{eq:symm:1}
\end{equation}
\begin{equation}
F(t+T/2)=-F(t) \quad \forall t\in\Re.
\label{eq:symm:2}
\end{equation}

Our main interest in this paper is to gain a deeper insight into
these type of systems exploring some quantitative and qualitative
aspects of its very rich dynamics. Specifically, we will consider
the same sawtooth potential as in Refs.~\cite{barhan94,goyhan00}
\begin{equation}
 U(x)=-\frac 1{2\pi} \left [ \sin (2\pi x)+\frac 14 \sin (4\pi x) \right ],
 \label{eq:uofx}
\end{equation}
which has a spatial period $\lambda=1$.
In addition, rather than using a sinusoidal driving force $F(t)$, in this work we will
consider a multifrequency time-periodic force given by
\begin{equation}
\label{pulse}
F(t)= \left \{ \begin{array}
{r@{\quad \mbox{for} \quad}l}
A & 0\le t < \frac T2 \\
-A& \frac T2 \le t < T, \\
\end{array}
\right .
\end{equation}
with $A$ being a constant. Since the potential (\ref{eq:uofx}) breaks
the spatial symmetry (\ref{eq:symm:1}), a directed current is possible,
even though the time symmetry (\ref{eq:symm:2}) is fulfilled.

The paper is organized as follows. In Section~\ref{sec:num}, we
perform a detailed numerical study of Eq.~(\ref{eq:eom}) for a
wide range of the driving parameter values $A$ and $\omega=2\pi/T$.
In particular, we present a colored phase-diagram $v$ vs. $A$ and $\omega$,
which is inspired by the figure provided by Prof.~Peter Talkner and
collaborators in \cite{goyhan00}. In order to explain some features of this
diagram we propose in Section~\ref{sec:ana} some simple theoretical considerations.
Finally, in the last section, we summarize our findings.

\section{Dynamical regimes and numerical evaluation of the current}
\label{sec:num}

\begin{figure}
\centerline{\includegraphics[width=6cm]{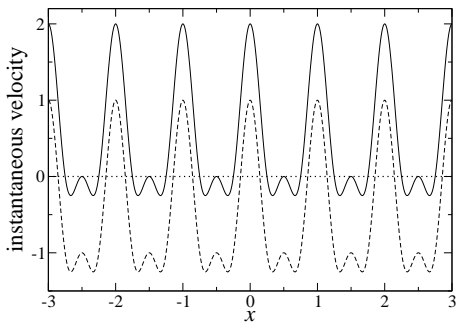}}
\caption{\label{fig:vofx} Plots of the instantaneous velocities $-U^\prime(x)+A$
(solid line) and  $-U^\prime(x)-A$ (dashed line) as a function of $x$  for a driving
 force with $A=0.5$. (All quantities in dimensionless
 units).}
\end{figure}

Regardless of the value of the driving frequency, there exist two critical amplitudes $A^*_{1}=3/4$ and
$A^*_{2}=3/2$ separating three different regimes.
For amplitudes $A \in [0,A^*_{1}]$, both potential states, $U(x)+Ax$ and $U(x)-Ax$, possess a periodic array of
equilibrium points (see Fig.~\ref{fig:vofx}). Since the inertial term $m \ddot{x}(t)$
is absent, the particle cannot cross these equilibrium points, and consequently,
it remains trapped between them, leading to a zero current. For $A \in (A^*_{1},A^*_{2}]$,
the potential state $U(x)+Ax$ retains its equilibrium points, while $U(x)-Ax$ does not have any,
that allowing non-bounding motion in the positive direction. Finally, when $A>A^*_{2}$,
neither of the potential states possesses equilibrium points, and the particle motion is not bounded in either direction.

\begin{figure}
\centerline{\includegraphics[height=9cm]{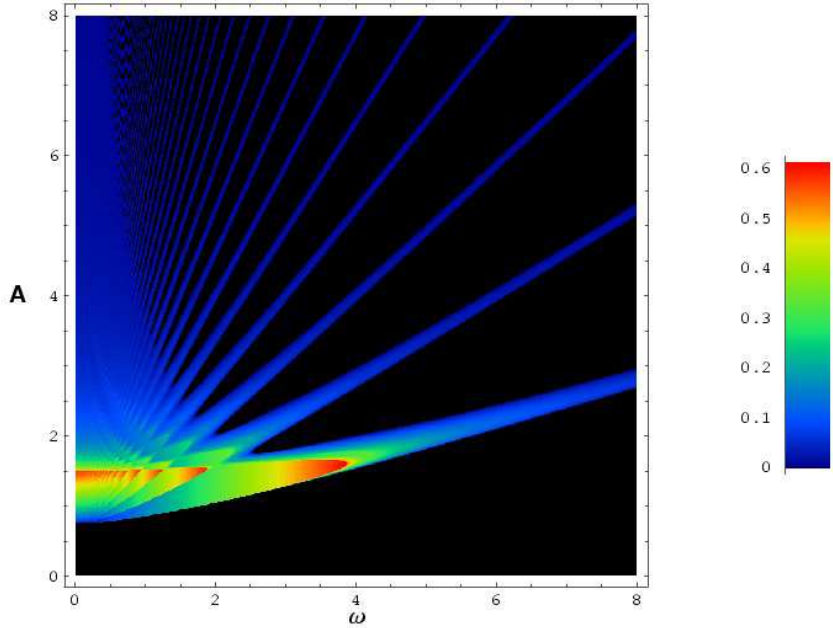}}
\caption{\label{fig:diagram} The "Talkner''-current phase diagram.
The drift velocity as a function of the driving parameters $A$ and
$\omega$ is represented using a color density plot. Black color has
been used for velocities less than $10^{-6}$. (All quantities in
dimensionless
 units)}
\end{figure}

\begin{figure}
\centerline{\includegraphics[height=6cm]{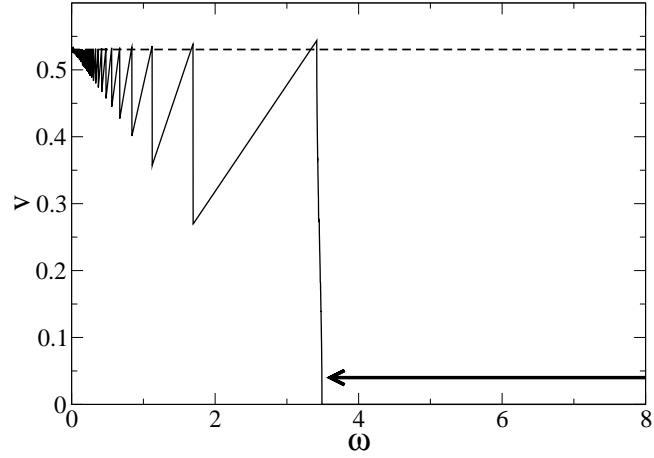}}
\caption{\label{fig:velOfw:1} Drift velocity $v$ (solid line) as a
  function of the driving frequency $\omega$ for $A=1.4< A^*_2$. The
  dashed line shows the adiabatic value. The
 arrows depicts the zero-velocity bands given by the theory in the text.
 (All quantities in dimensionless
 units.)}
\end{figure}

\begin{figure}
\centerline{\includegraphics[height=6cm]{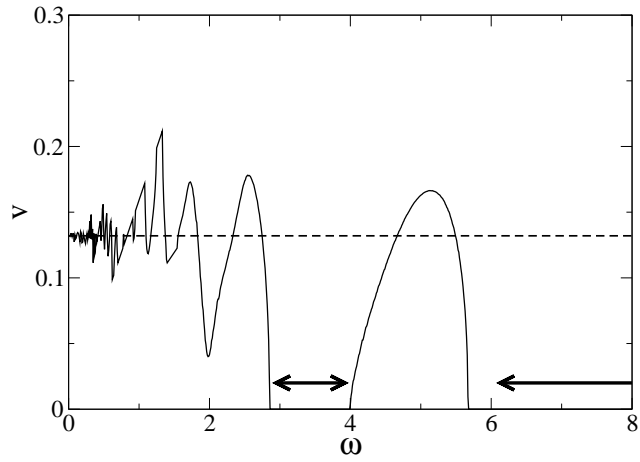}}
\caption{\label{fig:velOfw:2} The same as in Fig.~\ref{fig:velOfw:1} but for  $A=2> A^*_2$.}
\end{figure}

In order to go further in our analysis, we have resorted to a
numerical treatment of our model using the freely available
integrator RKSUITE \cite{rksuite}. Wide regions of parameter space
have been explored. In Fig.~\ref{fig:diagram} we have used a color
code to represent the drift velocity $v$ for a rectangular driving
force of varying amplitude $A$ and frequency $\omega$. The regions
in black correspond to bounded, time-periodic particle motion with
zero current. More precisely, the velocity in those regions is
smaller than $10^{-6}$. The diagram has a rich and interesting
structure. Regions of zero drift velocity are intermingled with
regions with finite values, giving rise to a finger-shaped
structure. Notice that for the parameter values considered, the
drift velocity is always positive or zero, showing the largest
magnitude in the intermediate region $A \in (A^*_{1},A^*_{2}]$. This
last feature could be understood by taking into account that in this
regime the motion in the positive direction is never compensated by
motion in the opposite direction.

In Fig.~\ref{fig:velOfw:1} we present a section of the diagram for a driving amplitude
in the intermediate regime  $A=1.4$. A series of peaks corresponding to the fingers in Fig.~\ref{fig:diagram} are observed.
Figure~\ref{fig:velOfw:2} shows a representative section for $A>A^*_2$. By contrast
with Fig.~\ref{fig:velOfw:1}, it presents an intermediate gap of particle localization.
If we further increase the value of $A$, more intermediate gaps appear, as can be seen in the phase diagram.

\section{Some theoretical results}
\label{sec:ana}
Even though the nonlinearity of the system precludes a complete and
detailed analytical solution of the problem, it is possible to explain
some features of the phase diagram by simple considerations.

\subsection{Proof of existence of regions displaying particle localization}

\begin{figure}
\centerline{\includegraphics[height=6cm,angle=0]{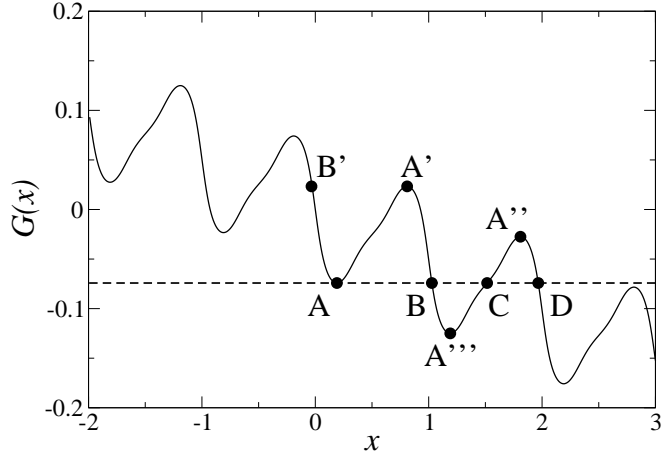}}
\caption{\label{fig:GofX} Determination of zero-velocity bands with
  $x(t+T)=x(t)$. Solid line depicts $G(x)$, defined in
  Eq.~(\ref{eq:GofX}), as a function of $x$ for the same value of $A$
  as in Fig.~\ref{fig:velOfw:2}.}
\end{figure}

Due to the truncation implicit in any numerical calculation, the simulations
reported above are not able to distinguish between a situation of exact
particle localization or a very small drift, lower than the tolerance we
chose to determine the black regions in Fig.\ref{fig:diagram}. Furthermore,
a simple explanation of the existence of such regions in the non-trivial regime $A>A^*_1$ would be desirable.

Let us assume that the particle starts at $x_0$ at the
beginning of a driving period. It then moves under a force $-U^\prime(x)+A$ for
half a period until it reaches the position $x_1> x_0$. Then the
driving force switches sign so the total force on the particle is now
$-U^\prime(x)-A$, which makes it (in general) move backwards up to a
position $x_0^\prime$ after another $T/2$. If $x_0^\prime=x_0$, the particle
has returned to the initial position, and consequently, the process is
repeated successively, leading to a drift velocity strictly equal to zero. For this situation to
happen the following equations must hold
\begin{eqnarray}
\frac T2&=&\int_{x_0}^{x_1}\frac{dx}{-U^\prime(x)+A} \label{eq:Tx0x1:1}\\
\frac T2&=&\int_{x_1}^{x_0}\frac{dx}{-U^\prime(x)-A}\label{eq:Tx0x1:2}.
\end{eqnarray}
Subtracting both equations we arrive to the condition
\begin{equation}
G(x_0)=G(x_1),
\end{equation}
where
\begin{equation}
G(x)=\int_0^x d\tilde{x} \frac{U^\prime(\tilde{x})}{
  A^2-U^\prime(\tilde{x})^2}.
\label{eq:GofX}
\end{equation}
Therefore, we can determine the set of pairs $(x_0,x_1)$ with zero
current for a given driving strength $A$ by plotting the function
$G(x)$ vs. $x$ (see Fig~\ref{fig:GofX}). The intersection of a
horizontal line with $G(x)$ in this plot provides the possible
values of the pair $(x_0,x_1)$. The period $T$ associated with the
pair is then given by Eq.~(\ref{eq:Tx0x1:1}) or
(\ref{eq:Tx0x1:2}). Since the drift velocity $v$ is independent of the
initial conditions \cite{adjmuk94}, it would be exactly zero for those driving parameter values $A$ and $T$.

In Fig.~\ref{fig:GofX}, we plot $G(x)$ for the same value of $A=2 (>A_2^*)$
as in Fig.~\ref{fig:velOfw:2}. A horizontal line crosses $G(x)$ at the
points A, B, C, and D, providing the coordinates $x_A, x_B, x_C$, and $x_D$. If
we choose $x_0=x_A$, and $x_1$ as any of the other points, we obtain
three pairs of points with driving periods $T_{AB}, T_{AC}$, and
$T_{AD}$. Since the integrand in Eq.~(\ref{eq:Tx0x1:1}) is always
positive, the further away $x_1$ is from $x_0$, the larger the
period, which implies $T_{AB}<T_{AC}<T_{AD}$.

Before we proceed in analyzing the plot in greater detail, let us
study the symmetries of $G(x)$. Since it is an integral of a
space-periodic function [see Eq.~(\ref{eq:GofX})] we have
\begin{equation}
G(x+1)=G(x)+\phi,
\label{eq:GofXsym:1}
\end{equation}
where $\phi=G(1)$. In addition, taking into account that $U^\prime(x)$
is an even function, necessarily
\begin{equation}
G(-x)=-G(x).
\label{eq:GofXsym:2}
\end{equation}
Property (\ref{eq:GofXsym:1}) leads to the fact that we only need to vary $x_0$
within a spatial interval of length unity, whereas property (\ref{eq:GofXsym:1}) combined
with (\ref{eq:GofXsym:2}) implies that $G(x)$ has an inversion center
at $x=n$, with $n$ any integer. The driving period $T$, when viewed as a function of
$x_1$ or $x_0$ [see Eq.~(\ref{eq:Tx0x1:1})], also obeys these properties.

Because of these symmetry properties, choosing the point A as $x_0$ in
Fig.~\ref{fig:GofX} is equivalent to choosing any of the points A{'},  A{'}{'}, and
A{'}{'}{'}, in the sense that they lead to the same driving periods.
Let us consider a horizontal line between points A and A{'}. As this line gets closer to A,
two of the intersection points approach each other, and consequently, the driving period associated to them tends to zero.
Therefore, $T=0$ ($\omega=\infty$) corresponds to particle localization. Moving
up the line continuously we obtain a whole band of driving periods
starting from zero up
to a maximum value given by the pair $(x_{B^\prime},x_{A^\prime})$ when
the line crosses A{'}. This pair is equivalent to $(x_{A},x_{B})$. Two more
intermediate periods in the band can be obtained from the pairs $(x_{B},x_{C})$ and $(x_C,x_D)$ in the figure.

The next value of the driving period obtained from
Fig.~\ref{fig:GofX} is the one given by the pair $(x_A,x_C)$. This pair marks the start of a
second band that ends at the maximum value given by
the pair $(x_A,x_D)$. Moving up the line up to A{'}{'} gives all the
intermediate values in the band.

Clearly, the first band is associated with movement within a spatial
interval less than unity (i.e., $0\le x_1-x_0<1$), whereas the
second band is related to displacements two times that distance
($1\le x_1-x_0<2$). In Fig.~\ref{fig:velOfw:2} we have indicated
with arrows the calculated frequency bands. It can be seen that they
do not cover the entire region with numerically evaluated zero
velocity. This is due to the fact that the above discussed mechanism
is just the simplest one leading to zero current. The drift velocity
can also vanish because the particle returns to its initial position
after two or more driving periods, instead of after the first one.

The same analysis can be carried out for a subthreshold driving
$A\in(A^*_1,A^*_2]$. In this case $G(x)$ presents singularities at the
equilibrium points, which prevents the particle from crossing those points
when $F(t)=-A$. This leads to a single band which is near the zero-current region observed in the
simulations, as shown in Fig.~\ref{fig:velOfw:1}.

\subsection{Adiabatic limit}

\begin{figure}
\centerline{\includegraphics[height=6cm]{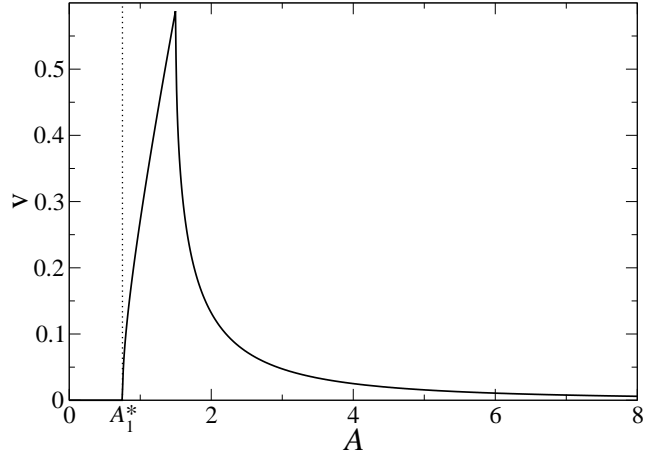}}
\caption{\label{fig:velOFom} Drift velocity $v$ (solid line) as a
  function of the driving amplitude $A$ in the adiabatic limit ($\omega\rightarrow 0$).
  The curve is indistinguishable from the numerical results obtained with $\omega=0.01$.
  (All quantities in dimensionless
 units.)}
\end{figure}

In this section we will obtain an analytical expression for the current
in the adiabatic limit $\omega \rightarrow 0$
\begin{equation}
v_{\mathrm{ad}}(A)=\lim_{\omega \rightarrow 0} v(A,\omega).
\end{equation}
Specifically, as proved later on, $v_{\mathrm{ad}}(A)$ is
given by the average value
\begin{equation}
\label{adiab}
v_{\mathrm{ad}}(A)=\frac{1}{2} \left[v_{+}(A)+ v_{-}(A)\right],
\end{equation}
where $v_{+}(A)$ and $v_{-}(A)$ are, respectively, the velocities in
the presence of the static forces $-U'(x)+A$ and
$-U'(x)-A$. Obviously, $v_{+}(A)=0$ for $A\in[0,A_{1}^*]$ and
$v_{-}(A)=0$ for $A\in[0,A_{2}^*]$, as the particle ends up being
trapped by an equilibrium point. For $A> A_{1}^*$,
\begin{equation}
v_{+}(A)=\frac{1}{\tau_{+}(A)},
\end{equation}
where
\begin{equation}
\label{timepos}
\tau_{+}(A)=\int_0^1
\frac{dx}{A-U'(x)}=\frac{2\sqrt{2A+\sqrt{-3+4A(1+A)}}}{\sqrt{(-1+2A)
(3+2A)(-3+4A)}}
\end{equation}
is the time taken for the particle to travel a distance equal to $1$
(the spatial period) in the presence of the static force
$-U'(x)+A$. Analogously, for $A> A_{2}^*$,
\begin{equation}
v_{-}(A)=\frac{-1}{\tau_{-}(A)},
\end{equation}
where
\begin{equation}
\label{timeneg}
\tau_{-}(A)=\int_0^1 \frac{dx}{A+U'(x)}=\frac{\sqrt{4A+2\sqrt{3+4A}}+
\sqrt{4A-2\sqrt{3+4A}}}{\sqrt{(-3+2A) (1+2A)(3+4A)}}
\end{equation}
is defined as $\tau_{+}(A)$ but replacing $-U'(x)+A$ by $-U'(x)-A$.

In order to prove Eq.~(\ref{adiab}), let us consider separately the
three regimes $A\in [0,A_{1}^*]$, $A\in (A_{1}^*,A_{2}^*]$, and $A\in
(A_{2}^*,\infty)$. For $A\in [0,A_{1}^*]$ the result in
Eq.~(\ref{adiab}) is trivial, since
$v_{\mathrm{ad}}(A)=v_{-}(A)=v_{+}(A)=0$. In the intermediate regime
$A\in (A_{1}^*,A_{2}^*]$, let us assume that the particle is
initially located at a minimum of the potential $U(x)+A x$ (as we have
mentioned before, the drift velocity does not depend on this particular
initial condition). If we choose the time period of $F(t)$ as $T=2 N
\tau_{+}(A)$, with $N=1$, $2$, $3$, $\dots$, then, during the first
half-period, $N \tau_{+}(A)$, the particle stays trapped at the
initial location. After the second half-period, $2N \tau_{+}(A)$, the
particle arrives to a new minimum of $U(x)+A x$ separated from the
initial location by a distance $N$. During the third half-period, $3N
\tau_{+}(A)$, the particle stays trapped at that minimum, and so on.
Thus, for a given $A\in (A_{1}^*,A_{2}^*]$, all the frequencies
\begin{equation}
\omega_N(A)=\frac{\pi}{N \tau_{+}(A)}
\label{eq:freq:1}
\end{equation}
lead to the same value of the drift velocity
\begin{equation}
\label{values}
v\left[A,\omega_N(A)\right]=\frac{N}{2 N \tau_{+}(A)}=\frac{1}{2
\tau_{+}(A)}=\frac{v_{+}(A)}{2}.
\end{equation}
Consequently, taking into account that $\lim_{N\rightarrow
\infty}\omega_N(A)=0$, it follows that
\begin{equation}
v_{\mathrm{ad}}(A)=\lim_{N\rightarrow \infty}
v\left[A,\omega_N(A)\right]=\frac{v_{+}(A)}{2}.
\end{equation}
This proves Eq.~(\ref{adiab}) for $A\in (A_{1}^*,A_{2}^*]$, since in
this regime $v_{-}(A)=0$.

Finally, for $A\in (A_2^*,\infty)$, it is easy to prove that
$\tau_{+}(A)/\tau_{-}(A)$ is a continuous strictly increasing function
of $A$ which takes values in the interval $(0,1)$. Let us assume
first that for a given $A\in (A_2^*,\infty)$ the ratio
$\tau_{+}(A)/\tau_{-}(A)$ is a rational number $p/q$, with
$p<q$. Then, if we choose the time period of $F(t)$ as $T=2 N q
\tau_{+}(A)=2 N p \tau_{-}(A)$, with $N=1$, $2$, $3$, $\dots$, the
particle travels a distance equal to $N(q-p)$ every time
period. Consequently, all the frequencies
\begin{equation}
\omega_N(A)=\frac{\pi}{N q \tau_{+}(A)}=\frac{\pi}{N p \tau_{-}(A)}
\label{eq:freq:2}
\end{equation}
lead to the same value of the drift velocity
\begin{equation}
v\left[A,\omega_N(A)\right]=\frac{N(q-p)}{2N q
\tau_{+}(A)}=\frac{1}{2\tau_{+}(A)}-\frac{1}{2\tau_{-}(A)}=
\frac{v_+(A)+v_-(A)}{2}.
\end{equation}
Taking into account once again that $\lim_{N\rightarrow
\infty}\omega_N(A)=0$, it follows that
\begin{equation}
v_{\mathrm{ad}}(A)=\lim_{N\rightarrow \infty}
v\left[A,\omega_N(A)\right]=\frac{v_{+}(A)+v_{-}(A)}{2}.
\end{equation}
Since any irrational number can be
approximated as closely as desired by a sequence of rational numbers,
Eq.~(\ref{adiab}) is also valid if $\tau_{+}(A)/\tau_{-}(A)$ is an
irrational number.

In Fig.~\ref{fig:velOFom} we represent the adiabatic drift velocity in Eq.~(\ref{adiab})
as a function of the driving amplitude $A$.  We have also plotted the numerical drift
velocity obtained for $\omega=0.01$. Both results are indistinguishable within the resolution of the plot.

Finally, from the above argument, another conclusion  follows. As
mentioned before, the adiabatic value $v_\mathrm{ad}(A)$ not only is
reached in the limit $\omega\rightarrow 0$, but also for the finite
frequencies given by Eqs.~(\ref{eq:freq:1}) and (\ref{eq:freq:2}).
This result becomes manifest in Figs.~\ref{fig:velOfw:1} and
\ref{fig:velOfw:2}, as an infinite number of intersections between
the horizontal line (adiabatic limit) and the drift velocity curve.
These frequencies are at the core (see the red regions in
Fig.~\ref{fig:diagram}) of the above discussed fingers. Furthermore,
they correspond to exact displacements of $n$ spatial periods every
driving period $T$. This fact, together with the interpretation
provided by our analysis of the zero-current bands, lead us to infer
that each finger corresponds to motion in which the particle
advances a distance of about $n$ spatial periods at those time
intervals when $F(t)=A$. We have checked this fact in the numerical
simulations.

\section{Conclusions}

To summarize, we have studied a one-dimensional deterministic
ratchet model in the overdamped regime driven by a rectangular
force. We have performed extensive numerical calculations, which
allowed us to present a colored phase diagram showing the drift
velocity as a function of a wide range of the driving parameter
values $A$ and $\omega$. In addition, we have studied the conditions
leading to bounded particle motion, obeying asymptotically periodic,
bounded motion, $x(t+T)=x(t)$. We have shown that the resulting
regions cover most of the phase space where there is no directed
current. Moreover, we have provided an explicit analytical
expression for the drift velocity in the adiabatic limit. Finally,
the theoretical considerations are of great help to rationalize the
finger-shaped structure shown in the diagram.\\

We (DC, JC-P, AA, MM) acknowledge the support of the Ministerio de
Educaci{\'o}n y Ciencia of Spain (FIS2005-02884) and the Junta de
Andaluc\'{\i}a. DC also acknowledges the Ministerio de Educaci{\'o}n
y Ciencia of Spain for a contract under the  Juan de la Cierva
program. The authors wish Professor Talkner many more enjoyable
years of doing trend-setting science and personal happiness.


\begin{thebibliography}{breitestes Label}
\bibitem{hanbar96}
P. H\"anggi and R. Bartussek, Lect. Notes Phys. \textbf{476}, 294
(1996).
\bibitem{astumian2002}
R.D. Astumian and P. H\"anggi, Physics Today \textbf{55}, (No. 11),
33 (2002).
\bibitem{linke02}
H. Linke, Appl Phys. A \textbf{75}, 167 (2002).
\bibitem{hanggi05}
P. H\"anggi, F. Marchesoni, and F. Nori, Ann. Phys. (Leipzig)
\textbf{14}, 51 (2005).
\bibitem{reimann02}
P. Reimann and P. H\"anggi, Appl. Phys. A \textbf{75},  169 (2002).
\bibitem{prostRMP}
F. J\"ulicher, A. Adjari, and J. Prost, Rev. Mod. Phys. \textbf{69},
1269 (1997).
\bibitem{junkis96}
P. Jung, J. G. Kissner, and P. H{\"a}nggi, Phys. Rev. Lett.
\textbf{76}, 3436 (1996).
\bibitem{mat00}
J. L. Mateos, Phys. Rev. Lett. \textbf{84}, 258 (2000).
\bibitem{sengupta04}
S. Sengupta, R. Guantes, S. Miret-Artes, and P. H\"anggi, Physica A
\textbf{338},  406 (2004).
\bibitem{ferlar05}
F. Family,  H. A. Larrondo, D. G. Zarlenga, and C. M. Arizmendi, J.
Phys. -- Condens Matter, \textbf{17}, S3719 (2005).

\bibitem{barhan94}
R. Bartussek, P. H{\"a}nggi, and J. G. Kissner, Europhys. Lett.
\textbf{28},  459 (1994).
\bibitem{sarlar99}
A. Sarmiento and H. Larralde, Phys. Rev. E \textbf{59}, 4878 (1999).
\bibitem{popari00}
M. N. Popescu, C. M. Arizmendi,  A. L. Salas-Brito, and F. Family,
Phys. Rev. Lett. \textbf{85}, 3321 (2000).

\bibitem{talkneritem}
M. Kostur, P. H\"anggi, P. Talkner, and J. L. Mateos, Phys. Rev. E
\textbf{72}, 036210 (2005).


\bibitem{adjmuk94}
A. Adjari, D. Mukamel, L. Peliti, and J. Prost, J. Phys.  I (France)
\textbf{4}, 1551 (1994).

\bibitem{goyhan00}
I. Goychuk and P. H\"anggi, Lect. Notes Phys. \textbf {557}, 7
(2000).

\bibitem{rksuite}
1991, http://www.netlib.org/ode/rksuite/

\end{thebibliography}
\end{document}